\documentclass[aps, pra, twocolumn, superscriptaddress, amsmath, showpacs, tightenlines, footinbib, longbibliography]{revtex4-2}
\usepackage[colorlinks, breaklinks, citecolor=blue, linkcolor=blue,  urlcolor={blue}]{hyperref}
\usepackage{breakurl}
\usepackage{amsfonts}
\usepackage{amsmath}
\usepackage{amssymb}
\usepackage{mathrsfs}
\usepackage{epsfig}
\usepackage{chngcntr}

\usepackage{bm}

\usepackage{colortbl}
\definecolor{mygray}{gray}{.9}
\definecolor{darkblue}{rgb}{1,1,.70}
\definecolor{lightblue}{rgb}{1,1,.90}

\allowdisplaybreaks[1]

\begin{document}

\title{Switching classical and quantum nonreciprocities with spinning photonics}

\author{Yonglin Xiang}
\thanks{These authors contributed equally to this work.}
\affiliation{Key Laboratory of Low-Dimensional Quantum Structures and Quantum Control of Ministry of Education, Department of Physics and Synergetic Innovation Center for Quantum  Effects and Applications, Hunan Normal University, Changsha 410081, China}

\author{Yunlan Zuo}
\thanks{These authors contributed equally to this work.}
\affiliation{Key Laboratory of Low-Dimensional Quantum Structures and Quantum Control of Ministry of Education, Department of Physics and Synergetic Innovation Center for Quantum  Effects and Applications, Hunan Normal University, Changsha 410081, China}

\author{Xun-Wei Xu}
\affiliation{Key Laboratory of Low-Dimensional Quantum Structures and Quantum Control of Ministry of Education, Department of Physics and Synergetic Innovation Center for Quantum Effects and Applications, Hunan Normal University, Changsha 410081, China}

\author{Ran Huang}
\email{ran.huang@riken.jp}
\affiliation{Theoretical Quantum Physics Laboratory, RIKEN Cluster for Pioneering Research, Wako-shi, Saitama 351-0198, Japan}

\author{Hui Jing}
\email{jinghui73@foxmail.com}
\affiliation{Key Laboratory of Low-Dimensional Quantum Structures and Quantum Control of Ministry of Education, Department of Physics and Synergetic Innovation Center for Quantum  Effects and Applications, Hunan Normal University, Changsha 410081, China}
\affiliation{Synergetic Innovation Academy for Quantum Science and Technology, Zhengzhou University of Light Industry, Zhengzhou 450002, China}
\date{\today}

\begin{abstract}
We study how to achieve, manipulate, and switch classical or quantum nonreciprocal effects of light with a spinning Kerr resonator. In particular, we show that even when there is no classical nonreciprocity (i.e., with the same mean number of photons for both clockwise and counterclockwise propagating modes), it is still possible to realize nonreciprocity of quantum correlations of photons in such a device. Also, by tuning the angular velocity and the optical backscattering strength, higher-order quantum nonreciprocity can appear, featuring qualitatively different third-order optical correlations, even in the absence of any nonreciprocity for both the mean photon number and its second-order correlations. The possibility to switch a single device between a classical isolator and a purely quantum directional system can provide more functions for nonreciprocal materials and new opportunities to realize novel quantum effects and applications, such as nonreciprocal multi-photon blockade, one-way photon bundles, and backaction-immune quantum communications.
\end{abstract}


\maketitle

\section{Introduction}
Optical nonreciprocity, featuring different responses of light when the input and output ports are interchanged, plays a key role in fundamental studies and applications of modern optics, such as directional laser engineering, invisible sensing, and backaction-immune optical communications~\cite{shoji2014Magnetooptical}. In recent years, without using any bulky magnetic material~\cite{adam2002Ferrite,dotsch2005Applications}, various ways have been demonstrated to create optical on-chip nonreciprocity, such as spatiotemporal modulation~\cite{yu2009Complete,kang2011Reconfigurable,estep2014Magneticfree,tzuang2014Nonreciprocal,koutserimpas2018Nonreciprocal,kittlaus2018Nonreciprocal,guo2019Nonreciprocal,wang2020Nonreciprocity}, optical nonlinearities~\cite{fan2012AllSilicon,shen2014Quantum,cao2017Experimental,xu2020Nonreciprocity,cao2020Reconfigurable,manipatruni2009Optical,dong2015Brillouinscatteringinduced,kim2015Nonreciprocala,shen2016Experimental,ruesink2016Nonreciprocity,hua2016Demonstration,bernier2017Nonreciprocal,miri2017Optical,shen2018Reconfigurable,rosariohamann2018Nonreciprocitya,sounas2018Nonreciprocity,mercierdelepinay2019Realization,xu2020Quantum,tang2021Broadintensityrange,shen2023Nonreciprocal}, non-Hermitian structures~\cite{ramezani2010Unidirectional,peng2014Paritya,chang2014Parity,huang2021Lossinduced}, quantum squeezing~\cite{tang2022Quantum}, and controllable motion of atoms or solid devices~\cite{wang2013Opticala,xia2018CavityFreea,zhang2018Thermalmotioninduceda,liang2020CollisionInduced,dong2021Alloptical,song2022Nonreciprocity,lu2017Optomechanically,maayani2018Flying,jiang2018Nonreciprocal,huang2018Nonreciprocal,li2019Nonreciprocal,wang2019Nonreciprocal,shen2020Nonreciprocal,jing2021Nonreciprocal,zhang2021Nonreciprocala,mirza2019Optical,jiao2020Nonreciprocal,
yang2020Nonreciprocal,ren2022Nonreciprocal,jiao2022Nonreciprocal,li2020Nonreciprocal,li2021Nonreciprocal,xu2021Nonreciprocal,shang2022Nonreciprocity,yang2022Nonreciprocal}. Particularly, by spinning a single device, it is possible to achieve nonreciprocal transmissions of light, sound, or thermal filed, without relying on any nonlinear medium~\cite{fleury2014Sound,khanikaev2015Topologically,xu2020Physical}, providing flexible new ways to achieve nanoparticle sensing~\cite{jing2018Nanoparticle,ahn2020Ultrasensitive,zhang2020Breaking}, optical gyroscopes~\cite{mao2022Experimental}, and quantum or topological directional control~\cite{dong2021Rotating,grinberg2020Robust,xu2020Tunable,zhu2021Inverse,xu2022Blackholeinspired}. Also, by further integrating with existing techniques of quantum nonlinear optics, purely quantum nonreciprocal effects can be achieved in such spinning systems, such as nonreciprocal photon blockade~\cite{huang2018Nonreciprocal,li2019Nonreciprocal,shen2020Nonreciprocal,wang2019Nonreciprocal,jing2021Nonreciprocal} and nonreciprocal quantum entanglement~\cite{jiao2020Nonreciprocal,yang2020Nonreciprocal,ren2022Nonreciprocal,jiao2022Nonreciprocal}. We note that, in a very recent experiment, quantum nonreciprocal correlations of photons were already observed in experiments using cavity atoms or an optical nonlinear system~\cite{yang2019Nonreciprocala,graf2022Nonreciprocity}. However, till now, in the absence of any classical nonreciprocity, the possibility of achieving one-way control of higher-order quantum correlations, has not yet been studied.

Here, in this work, we show how to achieve coherent switch of classical and quantum nonreciprocities of photons, and how to realize higher-order quantum nonreciprocity with a single spinning resonator. We find that, by tuning both the angular speed of the resonator and the optical backscattering strength, one can switch the functions of the device between a classical isolator and a purely quantum directional system. Also, a new class of higher-order quantum nonreciprocity, i.e., when both the mean photon numbers and the second-order correlations are reciprocal, the third-order correlation function is nonreciprocal. Particularly, we note that the backscattering due to material imperfections can induce a higher-order quantum nonreciprocal effect, in comparison with that in ideal devices. Our findings indicate a promising new way to achieve novel nonreciprocal effects, which is useful in realizing chiral quantum networks~\cite{gonzalez-ballestero2015Chiral,kimble2008quantum,lodahl2017Chiral,gangaraj2017Robust} and invisible sensing~\cite{fleury2015invisible,yang2015Invisible}.

The remainder of this paper is organized as follows. In Sec.~\ref{sec:2}, we introduce the physical system of a spinning Kerr resonator with a tapered fiber. In Sec.~\ref{sec:3}, we study the quantum and classical nonreciprocities in the ideal spinning resonator. In Sec.~\ref{sec:4}, we explore the quantum and classical nonreciprocities for systems with backscattering. In particular, revealing a novel higher-order quantum nonreciprocity. In Sec.~\ref{sec:5}, we give a summary.

\section{ Physical system\label{sec:2}}
\begin{figure*}[tbp]
\includegraphics[width=1\textwidth]{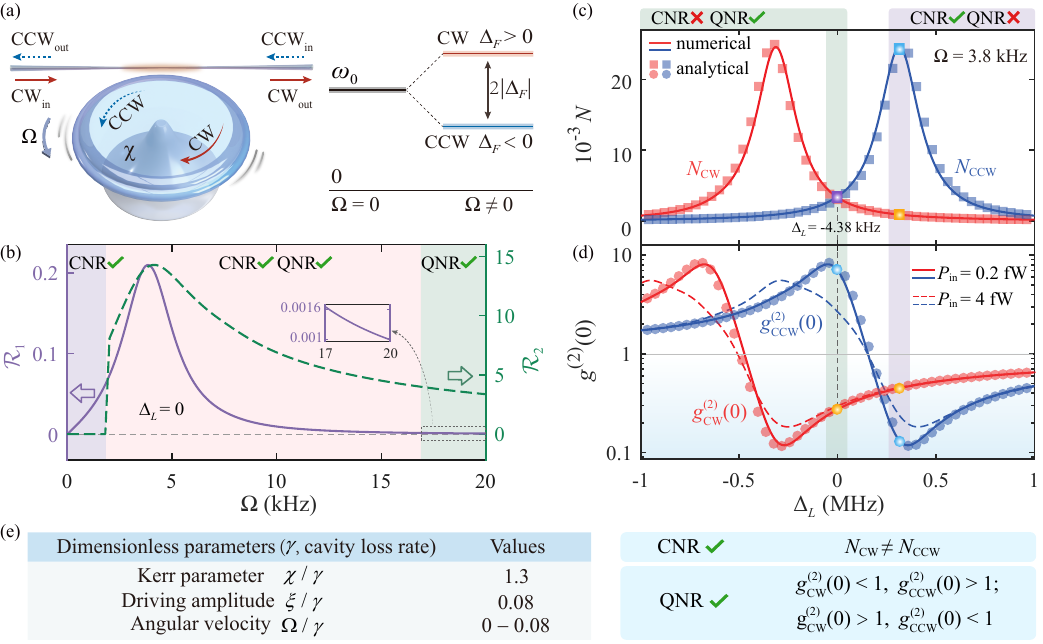}
\caption{Quantum and classical nonreciprocities. (a) The whispering-gallery-mode resonator with Kerr-type nonlinearity $\chi$ spinning at an angular velocity $\Omega$. By fixing the CCW rotation of the resonator, the $\Delta_{F}>0$ ($\Delta_{F}<0$) corresponds to the situation of driving the CW (CCW) mode.  (b) The classical (quantum) nonreciprocity ratio $\mathcal{R}_{1}$ ($\mathcal{R}_{2}$) versus the angular velocity $\Omega$ for $\Delta_{L}=0$, where the solid and dashed curves indicate $\mathcal{R}_{1}$ and $\mathcal{R}_{2}$, respectively. The inset shows that $\mathcal{R}_{1}\sim 0.0016$ when $\Omega=17\,\mathrm{kHz}$ (i.e., $\Omega/\gamma=0.07$). By further increasing $\Omega$, $\mathcal{R}_{1}$ gradually approaches 0. (c) The mean photon number $N$ and (d) the second-order correlation function $g^{(2)}(0)$ versus the optical detuning $\Delta_{L}$ for different input directions at $\Omega=3.8~\mathrm{kHz}$ (i.e., $\Omega/\gamma=0.016$). The markers
(squares, circles) and solid curves are analytical and numerical solutions at $P_\mathrm{in}$ = 0.2$\mathrm{~fW}$ (i.e., $\xi/\gamma=0.08$), respectively. When we take the parameter $P_\mathrm{in}$ = 4$\mathrm{~fW}$ (i.e., $\xi/\gamma=0.36$) in the experiment~\cite{schuster2008Nonlinear} (dashed curves), the quantum nonreciprocal feature still can be achieved. (e) The values and definitions of the main dimensionless parameters that we used in our calculations. The other parameters are given in the main text.}\label{Fig1}
\end{figure*}

As shown in Fig.~\ref{Fig1}(a), we consider a spinning Kerr resonator evanescently coupled with a tapered fiber, and each side of the fiber serves as both an input port and an output port. Depending on the input port, light is coupled to circulate in the resonator in either the clockwise (CW) or the counterclockwise (CCW) direction. Recently, $99.6\%$ optical isolation was realized experimentally by using a spinning resonator~\cite{maayani2018Flying}. In this experiment, the resonator is mounted on a turbine and rotated along its axis. The tapered fiber is made by heating and pulling a standard single-mode telecommunications fiber, which is stabilized at the height of several nanometers above the rapidly spinning resonator via the “self-adjustment” aerodynamic process. Here, for a resonator spinning along a fixed direction at an angular velocity $\Omega$, the resonance frequencies of the CW and CCW modes experience an opposite Sagnac-Fizeau shift, i.e., $\omega_0\to\omega_0+\Delta_{F}$, with~\cite{malykin2000Sagnac}
\begin{equation}
\Delta_{F}=\pm\frac{n_1r\Omega\omega_{0}}{c}\left(1-\frac{1}{n_1^{2}}-\frac{\lambda}{n_1}\frac{dn_1}{d\lambda}\right),
\end{equation}
where $\omega_{0}$ is the optical frequency of the nonspinning resonator, $c$ ($\lambda$) is the speed (wavelength) of light in a vacuum, $n_1$ and $r$ are the refractive index and radius of the resonator, respectively. The dispersion term $dn_1/d\lambda$, characterizing the relativistic origin of the Sagnac effect, is relatively small in typical materials (up to about 1\%)~\cite{maayani2018Flying,malykin2000Sagnac}. By spinning the resonator along the CCW direction, we have $\Delta_{F}>0$ ($\Delta_{F}<0$) for the case with the driven the CW (CCW) mode, i.e., $\omega_{\mathrm{cw},\mathrm{ccw}}=\omega_0\pm|\Delta_{F}|$, see Fig.~\ref{Fig1}(a).

In the frame rotating at the drive frequency $\omega_L$, the Hamiltonian of the system reads $(\hbar=1)$
\begin{equation}\label{H:r1}
\hat{H}_{1}=(\Delta_{L}+\Delta_{F})\hat{a}^{\dagger}\hat{a}+\chi\hat{a}^{\dagger}\hat{a}^{\dagger}\hat{a}\hat{a}+\xi(\hat{a}^{\dagger}+\hat{a}),
\end{equation}
where $\Delta_{L}=\omega_{0}-\omega_{L}$ is the optical detuning between the driving field and the cavity field, $\hat{a}$ ($\hat{a}^{\dagger}$) is the optical annihilation (creation) operator, $\chi=\hbar\omega_{0}^{2}c n_{2}/\left(n_{0}^{2}V_{\mathrm{eff}}\right)$ is the Kerr parameter, with the linear (nonlinear) refraction index $n_{0}\left(n_{2}\right)$, and the effective mode volume $V_{\text {eff }}$. The $\xi=\sqrt{\gamma P_{\mathrm{in}}/\left(\hbar\omega_{L}\right)}$ is  driving amplitude with the cavity loss rate $\gamma$ and the driving power $P_{\mathrm{in}}$.

The energy eigenstates of this system are the Fock states $\left|n\right\rangle$ $(n=0,1,2,...)$ with eigenenergies
\begin{equation}
E_{n}=n\Delta_{L}+n\Delta_{F}+(n^{2}-n)\chi,
\end{equation}
where $n$ is the cavity photon number. The positive and negative values of $\Delta_{F}$ describing upper or lower shifts of energy levels, respectively [Fig.~\ref{Fig1}(a)], and the values of $|\Delta_{F}|$ with an amount being proportional to $\Omega$. For the same input light, due to the opposite frequency shift for the counterpropagating modes, nonreciprocity can appear.

The experimentally accessible parameters we used here
are~\cite{zielinska2017self,schuster2008Nonlinear,vahala2003Optical,spillane2005Ultrahigh,pavlov2017Soliton,huet2016Millisecond}: $\lambda=1550\,\mathrm{nm}$, $Q=5\times10^9$, $V_{\text{eff}}=150\,\mu\mathrm{m}^3$, $n_2=2\times10^{-15}\,\mathrm{m}^2/\mathrm{W}$, $n_{0}=1.4$, $P_{\text{in }}$ = 0.2$\mathrm{~fW}$, and $r=30\,\mu\mathrm{m}$. In addition, we set the angular velocity $\Omega=3.8$ and $5.8~\mathrm{kHz}$, which are experimentally accessible parameters in Ref.~\cite{maayani2018Flying}. We  note that $V_{\text{eff }}$ is typically $10^2$--$10^4\,\mu\mathrm{m}^3$~\cite{vahala2003Optical,spillane2005Ultrahigh}, $Q$ is typically $10^9$--$10^{12}$~\cite{pavlov2017Soliton,huet2016Millisecond}. The Kerr coefficient can be $n_2=2\times10^{-15}\,\mathrm{m}^2/\mathrm{W}$ for materials with potassium titanyl phosphate~\cite{zielinska2017self}, i.e., $\chi/\gamma=1.3$. The driving power can reach $P_\mathrm{in}$ = 4$\mathrm{~fW}$ (i.e., $\xi/\gamma=0.36$) in the experiment~\cite{schuster2008Nonlinear}. Here, we have chosen the experimentally accessible value $\chi/\gamma=1.3$ and $\xi/\gamma=0.08$. Also we have confirmed that even for the values taken in the experiment~\cite{schuster2008Nonlinear}, i.e., $\xi/\gamma=0.36$, the quantum nonreciprocal feature still can be achieved, as shown in the Fig.~\ref{Fig1}(d) (dashed curves).

\section{Classical and quantum nonreciprocities\label{sec:3}}
The classical features of this work can be characterized by the mean photon number $N$:
\begin{equation}
N=\left\langle \hat{a}^{\dagger}\hat{a}\right\rangle.
\end{equation}
The condition of classical nonreciprocity (CNR) is
\begin{equation}\label{CNR}
N_{\mathrm{CW}} \neq N_{\mathrm{CCW}}.
\end{equation}
Note that in this work, we use subscripts $\mathrm{CW}$ and $\mathrm{CCW}$ to denote the cases with the driven the CW and CCW modes, respectively.

The quantum features can be characterized by the quantum correlation function $g^{(2)}(0)$:
\begin{equation}
g^{(2)}(0) = \frac{\left\langle \hat{a}^{\dagger2}\hat{a}^{2}\right\rangle }{\left\langle \hat{a}^{\dagger}\hat{a}\right\rangle ^{2}}.
\end{equation}
The conditions $g^{(2)}(0)<1$ and $g^{(2)}(0)>1$ characterize single-photon blockade and photon-induced tunneling~\cite{Faraon2008,Majumdar2012}, respectively, which are two distinct quantum effects with different photon-number statistics, i.e., photon antibunching and bunching~\cite{scully1997quantum,adam10}. Thus, we refer to the effect that photon antibunching in one direction and bunching in the other direction as quantum nonreciprocity. Here, the condition of quantum nonreciprocity (QNR) is given by
\[g_{\mathrm{CW}}^{(2)}(0)<1,\quad g_{\mathrm{CCW}}^{(2)}(0)>1,\]
or
\begin{equation}\label{QNR}
g_{\mathrm{CW}}^{(2)}(0)>1,\quad g_{\mathrm{CCW}}^{(2)}(0)<1.
\end{equation}


To better study the classical and quantum nonreciprocities of this system, we define the classical and quantum nonreciprocity ratios as $\mathcal{R}_{1}$ and $\mathcal{R}_{2}$, respectively, which are written as
\begin{align}\label{eq:R}
\mathcal{R}_{1} & = 10\log_{10}\frac{N_{\mathrm{CCW}}}{N_{\mathrm{CW}}},\nonumber \\
\mathcal{R}_{2} & = 10\log_{10}\frac{g_{\mathrm{CCW}}^{(2)}(0)}{g_{\mathrm{CW}}^{(2)}(0)}.
\end{align}
When $N_{\mathrm{CW}}$ and $N_{\mathrm{CCW}}$ do not satisfy the condition of nonreciprocity in Eq.~(\ref{CNR}), i.e., classical reciprocity, $\mathcal{R}_{1}=0$. Similarly, when both of $g_{\mathrm{CW}}^{(2)}(0)$ and $g_{\mathrm{CCW}}^{(2)}(0)$  are larger or smaller than 1, i.e., the same quantum effects with the identical photon-number statistics occur in both of two directions, we refer to this effect as quantum reciprocity. Thus, we set $\mathcal{R}_{2} = 0$ for this case.

According to the quantum-trajectory method~\cite{plenio1998quantumjump}, the optical decay can be included in the effective Hamiltonian
\begin{equation}\label{Eq:He1}
\hat{H}_{\mathrm{e1}}=\hat{H}_1-i\frac{\gamma}{2}\hat{a}^{\dagger}\hat{a},
\end{equation}
with $\gamma=\omega_{0}/Q$ is the cavity loss rate with the quality factor $Q$. Under the weak-driving condition ($\xi\ll\gamma$), the Hilbert space can be truncated to $n=2$. The state of this system can be expressed as
\begin{equation}\label{Eq:phi1}
|\varphi(t)\rangle=\sum_{n=0}^{2}C_{n}(t)|n\rangle,
\end{equation}
with probability amplitudes $C_{n}$. Based on the effective Hamiltonian in Eq.~(\ref{Eq:He1}), and the wave function in Eq.~(\ref{Eq:phi1}), we can obtain the following equations of motion for the probability amplitudes $C_{n}(t)$:
\begin{align}
\dot{C}_{0}(t) & =-iE_0C_{0}(t)-i\xi C_{1}(t),\nonumber \\
\dot{C}_{1}(t) & =-i\left(E_1-i\frac{\gamma}{2}\right)C_{1}(t)-i\xi C_{0}(t)-i\xi\sqrt{2}C_{2}(t),\nonumber\\
\dot{C}_{2}(t) & =-i(E_2-i\gamma)C_{2}(t)-i\xi\sqrt{2}C_{1}(t),
\end{align}
where $\ensuremath{E_0=0}$, $E_1=\Delta_{L}+\Delta_{F}$,
$E_2=2(\Delta_{L}+\Delta_{F})+2\chi$. In the weak-driving case,  these probability amplitudes have the following approximation expressions: $C_{0}\sim1$, $C_{1}\sim\xi/\gamma$,
and $C_{2}\sim\xi^{2}/\gamma^{2}$. According to perturbation method~\cite{carmichael1991Quantum} to solve the above equations, we can obtain the probability amplitudes as
\begin{align}\label{c123}
\dot{C}_{0}(t) & =-iE_0C_{0}(t),\nonumber \\
\dot{C}_{1}(t) & =-i\left(E_1-i\frac{\gamma}{2}\right)C_{1}(t)-i\xi C_{0}(t),\\
\dot{C}_{2}(t) & =-i(E_2-i\gamma)C_{2}(t)-i\xi\sqrt{2}C_{1}(t).\nonumber
\end{align}
In an initially empty cavity, the initial conditions can be set as:
\begin{align}
C_{0}(0) & =C_{0}(0),\nonumber \\
C_{1}(0) & =C_{2}(0)=0.
\end{align}
Therefore, we can obtain the solution of the zero-photon amplitude:
\begin{equation}\label{C0t}
C_{0}(t)=C_{0}(0)\exp\left(-iE_0t\right).
\end{equation}
We introduce the slow-varying amplitudes to solve this equation:
\begin{align}
& C_{1}(t)=c_{1}(t)\exp\left[-i(E_1-i\frac{\gamma}{2})t\right],~~
C_{1}(0)=c_{1}(0),\nonumber \\
& C_{2}(t)=c_{2}(t)\exp\left[-i(E_2-i\gamma)t\right],\quad ~ C_{2}(0)=c_{2}(0).
\end{align}
Then, based on the solution of the zero-photon amplitude in Eq.~(\ref{C0t}) and the above equations, we can obtain the solutions of the equations of motion for the probability amplitudes
\begin{align}
C_{0}(t) & =C_{0}(0)\mathcal{A}_{0},\nonumber \\
C_{1}(t) & =-\xi C_{0}(0)\left(\mathcal{A}_{0}-\mathcal{A}_{1}\right)/\left(E_1-i\frac{\gamma}{2}\right),\nonumber \\
C_{2}(t) & =\sqrt{2}\xi^{2}C_{0}(0)\left(\mathcal{B}_{0}-\mathcal{B}_{1}\right)/\left(E_1-i\frac{\gamma}{2}\right),
\end{align}
where
\begin{align}
 & \mathcal{A}_{0}=\exp\left(-iE_0t\right),\nonumber \\
 & \mathcal{A}_{1}=\exp\left[-i\left(E_1-i\gamma/2\right)t\right],\nonumber \\
 & \mathcal{A_{\mathrm{2}}}=\exp\left[-i\left(E_2-i\gamma\right)t\right],\nonumber \\
 & \mathcal{B}_{0}=\left(\mathcal{A}_{0}-\mathcal{A}_{2}\right)/\left(E_2-E_0-i\gamma\right),\nonumber \\
 & \mathcal{B}_{1}=\left(\mathcal{A}_{1}-\mathcal{A}_{2}\right)/\left(E_2-E_1-i\gamma/2\right),
\end{align}
and for the infinite-time limit, we have $\exp(-At)\to0$ $(t\to\infty)$, then the solutions should be
\begin{align}
C_{0} & \equiv C_{0}(\infty)=1,\nonumber \\
C_{1} & \equiv C_{1}(\infty)=\frac{-\xi}{\left(E_1-i\gamma/2\right)},\nonumber \\
C_{2} & \equiv C_{2}(\infty)=\frac{-\sqrt{2}\xi C_{1}}{\left(E_2-i\gamma\right)}.
\end{align}
According to the normalized coefficient of the state
\begin{equation}
\mathcal{M}=1+\left|C_{1}\right|^{2}+\left|C_{2}\right|^{2},
\end{equation}
we can get the probabilities of finding $n$ photons in the resonator as
\begin{equation}\label{Pm}
P_{n}=\frac{\left|C_{n}\right|^2}{\mathcal{M}}.
\end{equation}
The mean photon number is denoted by $N$, and can be obtained from the above probability distribution as
\begin{equation}
N=\left\langle \hat{a}^{\dagger}\hat{a}\right\rangle =\sum_{n=0}^{2}nP_{n}.
\end{equation}
For $\Delta_{L}=0$, the mean photon number becomes
\begin{align}
N=&{\frac{\xi^{4}}{\chi\gamma^{2}\Delta_{F}+4\chi\Delta_{F}^{3}+2\left(\Delta_{F}^{2}+\chi^{2}+\gamma^{2}/4\right)\left(\Delta_{F}^{2}+\gamma^{2}/4\right)}}\nonumber \\
&+\frac{\xi^{2}}{\Delta_{F}^{2}+\gamma^{2}/4}.
\end{align}
The origin of classical nonreciprocity can be understood from the terms of the Sagnac-Fizeau shift ($\propto\Delta_{F}$) and the cubic terms of it ($\propto\Delta_{F}^3$).

The equal-time (namely zero-time-delay) second-order correlation function is written as
\begin{equation}\label{g2ana}
g^{(2)}(0) =\frac{\langle\hat{a}^{\dagger2}\hat{a}^{2}\rangle}{\langle\hat{a}^{\dagger}\hat{a}\rangle^{2}}\simeq
\frac{(\Delta_{L}+\Delta_F)^2+\gamma^2/4}{(\Delta_{L}+\Delta_{F}+\chi)^2+\gamma^2/4}.
\end{equation}
For $\Delta_{L}=0$, photon antibunching [$g^{(2)}(0)<1$] and bunching [$g^{(2)}(0)>1$] occur under the conditions of $\Delta_{F}>-\chi/2$ and $\Delta_{F}<-\chi/2$, respectively. Therefore, the condition of quantum nonreciprocity, $\mathcal{R}_2\neq0$, is given by
\begin{equation}
\Omega>\frac{\chi}{2\alpha},\qquad \alpha=\frac{n_1r\omega_0}{c}\left(1-\frac{1}{n_1^2}-\frac{\lambda}{n_1}\frac{dn_1}{d\lambda}\right).
\end{equation}
This result is in agreement with the results of Fig.~\ref{Fig1}(b), i.e., the transition between quantum reciprocity and nonreciprocity occurs at $\Omega\approx 1.93\,\mathrm{kHz}$.

To obtain more exact results, we numerically study the full quantum dynamics of the system by solving the master equation~\cite{johansson2013QuTiP,johansson2012QuTiP}
\begin{equation}\label{eq:master}
\dot{\hat{\rho}}=\frac{i}{\hbar}[\hat{\rho}, \hat{H}_1]+\frac{\gamma}{2}\left(2 \hat{a} \hat{\rho} \hat{a}^{\dagger}-\hat{a}^{\dagger} \hat{a} \hat{\rho}-\hat{\rho} \hat{a}^{\dagger} \hat{a}\right),
\end{equation}
where $\hat{\rho}$ is the normalized density matrix of the system. The photon-number probability is $P_n=\langle n |\hat{\rho}_\text{ss}|n\rangle$, with the steady-state solutions $\hat{\rho}_\text{ss}$ of the master equation.

Figure \ref{Fig1}(b) shows the switching between CNR and QNR when the optical detuning $\Delta_{L}=0$. For a non-spinning resonator ($\Omega=0$), both classical and quantum effects are reciprocal at this point, i.e., $\mathcal{R}_{1}=0$ and $\mathcal{R}_{2}=0$. When the angular velocity $\Omega$ below $1.93\,\mathrm{kHz}$, the classical nonreciprocity appears ($\mathcal{R}_{1}>0$), since the rotation-induced Sagnac effect breaks the degeneracy of the CW and CCW modes leading to a mode splitting, which makes the light transparent in one direction while opaque in the other, i.e., the one-way transmission of the light~\cite{maayani2018Flying}. At the same time, the quantum nonreciprocity ratio $\mathcal{R}_{2}$ is always equal to $0$, the quantum effect is reciprocal.

When $\ensuremath{\Omega=[1.93\,\mathrm{kHz},17\,\mathrm{kHz}]}$, the quantum nonreciprocity ($\mathcal{R}_{2}>0$) emerges due to the interplay of both rotation-induced Sagnac effect and the nonlinearity-induced anharmonicity. The Kerr nonlinearity leads to an anharmonic energy space, which can be further shifted by Sagnac effect. When the rotation speed is large enough, the energy level of the system satisfies the conditions for photon antibunching in one direction and bunching in the other direction~\cite{huang2018Nonreciprocal}.

When the $\Omega$ exceeds $17\,\mathrm{kHz}$, $\mathcal{R}_{1}$ gradually approaches 0, while the quantum nonreciprocity still remains ($\mathcal{R}_{2}>0$), this means that quantum nonreciprocity exists even when there is no classical nonreciprocity. We note that quantum nonreciprocity can exist independently of classical nonreciprocity, as shown in the recent experiment~\cite{graf2022Nonreciprocity}. The fundamental reason is that classical and quantum nonreciprocities are two essentially different concepts, which are defined via the mean-photon number and quantum fluctuation. With optical nonlinearity, quantum fluctuations can be different with the same mean-photon numbers~\cite{graf2022Nonreciprocity}. With such a device, the different nonreciprocities can be tuned by tuning the angular velocity.

In addition, we note that the switching between classical nonreciprocity and pure quantum nonreciprocity can also be achieved by tuning the optical detuning $\Delta_{L}$ [Figs.~\ref{Fig1}(c) and \ref{Fig1}(d)]. As an illustration, for a spinning cavity, by driving the CW (CCW) mode, we have $\Delta_{F}>0$ ($\Delta_{F}<0$), thus, leading to quantum nonreciprocity at $\Delta_{L}=-4.38\,\mathrm{kHz}$, i.e., $g_\mathrm{CW}^{(2)}(0)\sim0.28$, $g_\mathrm{CCW}^{(2)}(0)\sim7$. At this point, $N_\mathrm{cw}=N_{\mathrm{ccw}}$, the classical effect is reciprocal. When the maximum difference of mean photon numbers by driving the setup from the right and left sides is generated, i.e., $N_\mathrm{CW}\sim0.001$ and $N_\mathrm{CCW}\sim0.0245$. This is a clear signature of classical nonreciprocity. At the same time, we have quantum reciprocity, i.e., $g_\mathrm{CW}^{(2)}(0)\sim0.45$ and $g_\mathrm{CCW}^{(2)}(0)\sim0.12$.

These results show that a single device switching between a classical isolator and a quantum one-way device can be achieved by adjusting multiple degrees of freedom~\cite{jiao2020Nonreciprocal}.

\section{Quantum and classical nonreciprocities with backscattering\label{sec:4}}

\begin{figure*}
\centering
\includegraphics[width=1\textwidth]{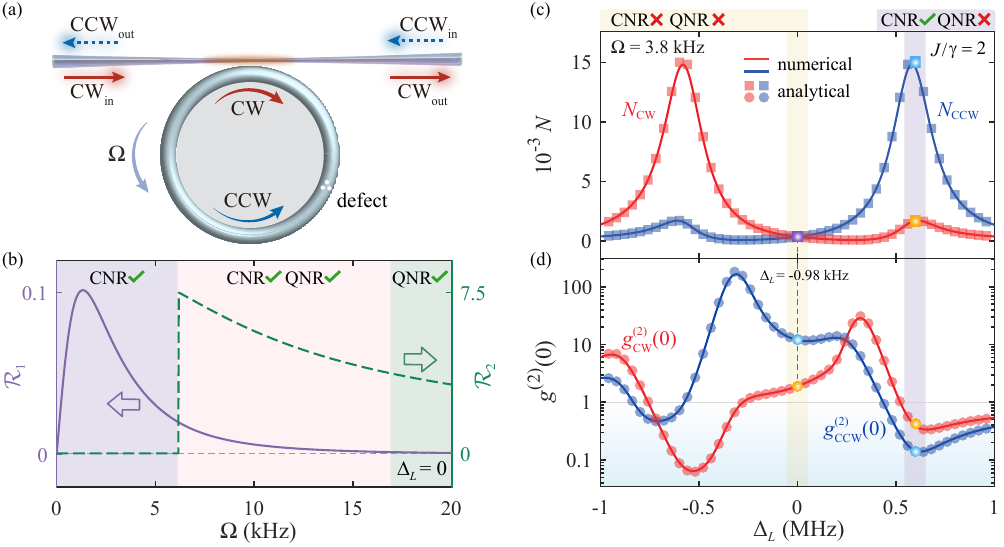}
\caption{Quantum and classical nonreciprocities with backscattering. (a) The spinning Kerr resonator with defect-induced backscattering. The CW and CCW modes are coupled via backscattering with the strength $J$. (b) The classical (quantum) nonreciprocity ratio $\mathcal{R}_{1}$ ($\mathcal{R}_{2}$) versus the angular velocity $\Omega$ for $\Delta_{L}=0$, where the solid (dashed) curve indicates $\mathcal{R}_{1}$ ($\mathcal{R}_{2}$). (c) The mean photon number $N$ and (d) the second-order correlation function $g^{(2)}(0)$ as functions of $\Delta_{L}$ for different input directions at $\Omega=3.8~\mathrm{kHz}$ (i.e., $\Omega/\gamma=0.016$). The markers (squares, circles) and solid curves are analytical and numerical solutions, respectively. We note that the backscattering rate can reach $J/\gamma=15$ in the experiment~\cite{zhu2010Onchip}, and the mode splitting induced by backscattering is easily experimentally observable for $J/\gamma>1/2$~\cite{kim2019Dynamic}. Here, the parameter of backscattering rate is chosen as $J/\gamma=2$, the other parameters are the same as those in Fig.~\ref{Fig1}.}
\label{Fig2}
\end{figure*}

Now, we further extend our present study to a  more generalized situation. In practice, the imperfections of devices, such as surface roughness or material defect, can cause optical backscattering, as shown in Fig.~\ref{Fig2}(a). Thus, we discuss the role of backscattering in quantum and classical nonreciprocities. For this aim, we introduce backscattering, as described by the coupling strength $J$ between the CW and CCW modes. The system’s Hamiltonian, given in Eq.~(\ref{H:r1}), is transformed to
\begin{align}\label{eq:H2}
\hat{H}_{2} = & \underset{j=1,2}{\sum}\Delta_{j}\hat{a}_{j}^{\dagger}\hat{a}_{j}+\underset{j=1,2}{\sum}\chi\hat{a}_{j}^{\dagger}\hat{a}_{j}^{\dagger}\hat{a}_{j}\hat{a}_{j} \nonumber\\
&+2\chi\hat{a}_{1}^{\dagger}\hat{a}_{1}\hat{a}_{2}^{\dagger}\hat{a}_{2}+J\left(\hat{a}_{1}^{\dagger}\hat{a}_{2}+\hat{a}_{2}^{\dagger}\hat{a}_{1}\right)
\nonumber\\
&+\xi(\hat{a}_{1}^{\dagger}+\hat{a}_{1}),
\end{align}
where $\hat{a}_1(\hat{a}_1^{\dagger})$ and $\hat{a}_2(\hat{a}_2^{\dagger})$ are the annihilation (creation) operators of the CW and CCW modes, respectively. And $\Delta_{j}=\Delta_{L}\pm|\Delta_F|(j=1,2)$, $2\chi\hat{a}_{1}^{\dagger}\hat{a}_{1}\hat{a}_{2}^{\dagger}\hat{a}_{2}$ is the cross-Kerr interaction~\cite{gong2009Effective,cao2020Reconfigurable,cao2017Experimental,heikkila2014Enhancing} between the CW and CCW modes.

Optical decay can be included in the effective Hamiltonian~\cite{plenio1998quantumjump},
\begin{equation}
\hat{H}_{\text{e2}}=\hat{H}_2-\sum_{j=1,2}i\gamma\hat{a}_{j}^{\dagger}\hat{a}_{j}/2.
\end{equation}
In the weak-driving case ($\xi\ll\gamma$), the Hilbert space of the system can be restricted to a subspace with few photons. Here, in order to calculate the expressions of the second-order and third-order correlation functions, this Hilbert space is truncated to a subspace with three-photon numbers, i.e., $N=m+n=3$, the wave function of the system can be expressed as
\begin{equation}
|\psi(t)\rangle=\sum_{N=0}^{3}\sum_{m=0}^{N}C_{m,N-m}|m,N-m\rangle,
\end{equation}
where $C_{mn}$ are probability amplitudes corresponding to state $|m,n\rangle$. By solving the Schrödinger equation
\begin{equation}
i|\dot{\psi}(t)\rangle=\hat{H}_{\mathrm{e}2}|\psi(t)\rangle,
\end{equation}
we can obtain the above probability amplitudes $C_{mn}$. When a weak-driving field is applied to the cavity, it may excite few photons in the cavity. Therefore, we can approximate the probability amplitudes of the excitations as $C_{m,N-m}\sim (\xi/\gamma)^{N}$. Here, we use the perturbation method~\cite{carmichael1991Quantum} to solve the equations of motion for the probability amplitudes $C_{m,N-m}(t)$. Then, we have the equations of motion for probability amplitudes
\begin{align}
\dot{iC}_{00}(t)= & 0,\nonumber\\
i\dot{C}_{01}(t)= & \Delta_{4}C_{01}(t)+JC_{10}(t),\nonumber\\
i\dot{C}_{10}(t)= & \Delta_{3}C_{10}(t)+JC_{01}(t)+\xi C_{00}(t),\nonumber\\
i\dot{C}_{02}(t)= & 2\Delta_{6}C_{02}(t)+\sqrt{2}JC_{11}(t),\nonumber\\
i\dot{C}_{11}(t)= & (\Delta_{5}+\Delta_{6})C_{11}(t)+\sqrt{2}JC_{20}(t)+\sqrt{2}JC_{02}(t)\nonumber\\
 & +\xi C_{01}(t),\nonumber\\
i\dot{C}_{20}(t)= & 2\Delta_{5}C_{20}(t)+\sqrt{2}JC_{11}(t)+\sqrt{2}\xi C_{10}(t),\nonumber\\
i\dot{C}_{03}(t)= & 3\Delta_{8}C_{03}(t)+\sqrt{3}JC_{12}(t),\nonumber\\
i\dot{C}_{12}(t)= & (\Delta_{7}+2\Delta_{8})C_{12}(t)+2JC_{21}(t)+\sqrt{3}JC_{03}(t)\nonumber\\
 & +\xi C_{02}(t),\nonumber\\
i\dot{C}_{21}(t)= & (2\Delta_{7}+\Delta_{8})C_{21}(t)+\sqrt{3}JC_{30}(t)+2JC_{12}(t)\nonumber\\
 & +\sqrt{2}\xi C_{11}(t),\nonumber\\
i\dot{C}_{30}(t)= & 3\Delta_{7}C_{30}(t)+\sqrt{3}JC_{21}(t)+\sqrt{3}\xi C_{20}(t),
\end{align}
where
\begin{align}\Delta_{3} & =\Delta_{1}-i\gamma/2,\qquad\Delta_{4}=\Delta_{2}-i\gamma/2,\nonumber\\
\Delta_{5} & =\Delta_{3}+\chi,\qquad\quad~\Delta_{6}=\Delta_{4}+\chi,\nonumber\\
\Delta_{7} & =\Delta_{5}+\chi,\qquad\quad~\Delta_{8}=\Delta_{6}+\chi.
\end{align}
By considering the initial condition $C_{00}(0)=1$ and setting $\dot{C}_{mn}(t)=0$, we can obtain the steady-state solutions of the probability amplitudes
\begin{align}
C_{10} & =\frac{\xi\Delta_{4}}{\eta_{1}},\qquad\thinspace\thinspace\qquad C_{01}=\frac{-\xi J}{\eta_{1}},\nonumber\\
C_{02} & =\frac{J^{2}\xi^{2}\sigma_{1}}{\sqrt{2}\eta_{1}\eta_{2}\sigma_{2}},\qquad C_{11}=-\frac{J\xi^{2}\Delta_{6}\sigma_{1}}{\eta_{1}\eta_{2}},\nonumber\\
C_{20} & =\frac{\xi^{2}\left(\Delta_{4}\Delta_{6}/\sigma_{2}+J^{2}\chi\right)}{\sqrt{2}\eta_{1}\eta_{2}\sigma_{2}},\nonumber\\
C_{03} & =-\frac{J^{3}\xi^{3}\Gamma_{4}}{\sqrt{6}\sigma_{2}\mu\eta_{1}\eta_{2}},\thinspace C_{12}=\frac{J^{2}\xi^{3}\Delta_{8}\Gamma_{4}}{\sqrt{2}\sigma_{2}\mu\eta_{1}\eta_{2}},\nonumber\\
C_{21} & =\frac{J\xi^{3}\left[\Gamma_{3}-\Delta_{3}\eta_{3}\left(\Delta_{3}+4\chi\right)\right]}{\sqrt{2}\sigma_{2}\mu\eta_{1}\eta_{2}},\nonumber\\
C_{30} & =\frac{\xi^{3}\left(\eta_{3}\Gamma_{1}-\Delta_{8}\Gamma_{2}\right)}{\sqrt{6}\sigma_{2}\mu\eta_{1}\eta_{2}},
\end{align}
with
\begin{align}\label{eq:tau} & \sigma_{1}=\Delta_{4}+\Delta_{5},\qquad\sigma_{2}=\Delta_{5}+\Delta_{6},\nonumber\\
 & \sigma_{3}=\Delta_{7}+\Delta_{8},\qquad\zeta=\sigma_{2}^{2}+\sigma_{3}\Delta_{7}-4J^{2},\nonumber\\
 & \eta_{1}=J^{2}-\Delta_{3}\Delta_{4},\quad\eta_{2}=J^{2}-\Delta_{5}\Delta_{6},\nonumber\\
 & \eta_{3}=J^{2}-\Delta_{7}\Delta_{8},\quad\mu=\eta_{3}\left(\eta_{3}-2\sigma_{3}^{2}\right),\nonumber\\
 & \Gamma_{1}=\left(\sigma_{3}+\Delta_{7}\right)(J^{2}\chi+\sigma_{2}\Delta_{4}\Delta_{6}),\nonumber\\
 & \Gamma_{2}=J^{2}\left[2\sigma_{1}\left(J^{2}+2\sigma_{3}\Delta_{6}\right)-\chi\zeta\right]+\sigma_{2}\Delta_{4}\Delta_{6}\zeta,\nonumber\\
 & \Gamma_{3}=\left(2\Delta_{8}^{2}-\eta_{3}\right)\left[J^{2}\chi+\Delta_{6}\left(\sigma_{2}\Delta_{4}+2\sigma_{1}\Delta_{7}\right)\right],\nonumber\\
 & \Gamma_{4}=\sigma_{1}\left(\eta_{3}-4\Delta_{6}\Delta_{7}\right)-2\left(J^{2}\chi+\sigma_{2}\Delta_{4}\Delta_{6}\right).
\end{align}

\begin{figure*}
\centering
\includegraphics[width=1\textwidth]{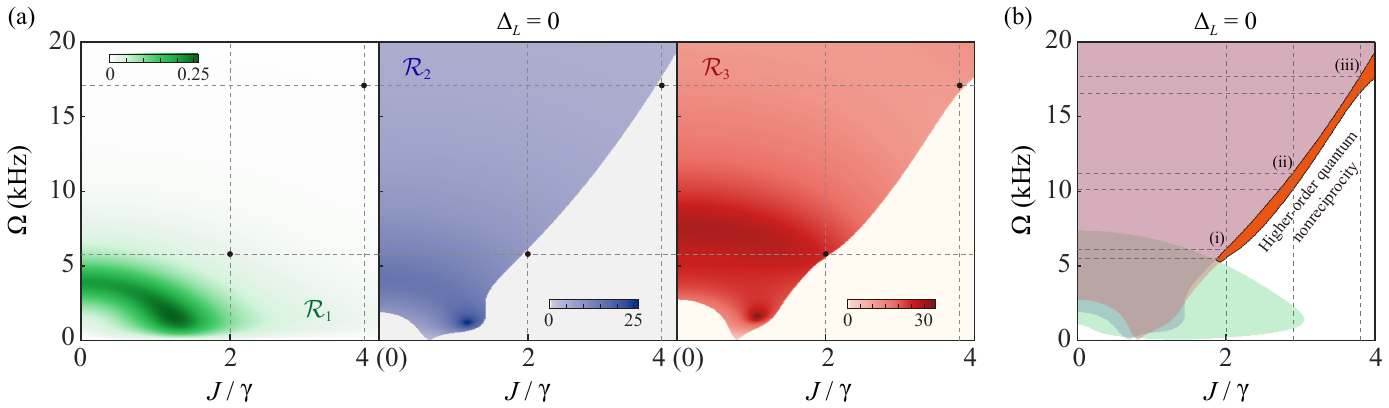}
\caption{(a) The classical nonreciprocity ratio $\mathcal{R}_1$, the quantum nonreciprocity ratio $\mathcal{R}_2$, and the higher-order quantum nonreciprocity ratio $\mathcal{R}_3$ obtained as function of backscattering rate $J$ and angular velocity $\Omega$. (b) The higher-order quantum nonreciprocity versus $J$ and $\Omega$. The other parameters are the same as those in Fig.~\ref{Fig1}.}\label{Fig3}
\end{figure*}

The probabilities of finding $m$ photons in the CW and $n$ photons in the CCW modes are given by
\begin{equation}
P_{mn}=\frac{\left|C_{mn}\right|^{2}}{\mathcal{M}},
\end{equation}
with the normalization coefficient
\begin{equation}
\mathcal{M}=\sum_{N=0}^{3} \sum_{m=0}^{N}\left|C_{m n}\right|^{2}.
\end{equation}
The mean photon number is expressed
\begin{align}
N&=\langle \hat{a}_{1}^{\dagger}\hat{a}_{1}\rangle\simeq P_{10}+P_{11}+2P_{20}\nonumber\\
&=\left|\frac{\xi\Delta_{4}}{\eta_{1}}\right|^{2}+\left|-\frac{J\xi^{2}\Delta_{6}\sigma_{1}}{\eta_{1}\eta_{2}}\right|^{2}+2\left|\frac{\xi^{2}\left(\Delta_{4}\Delta_{6}/\sigma_{2}+J^{2}\chi\right)}{\sqrt{2}\eta_{1}\eta_{2}\sigma_{2}}\right|^{2}.
\end{align}
Similarly, for $\Delta_{L}=0$, the Sagnac-Fizeau shift and the odd power of it, which are included in $P_{11}$ and $P_{20}$, lead to the classical nonreciprocity with a resonant drive, as shown in Fig.~\ref{Fig2}(b).\\
The second-order correlation function is written as
\begin{equation}\label{eq:g2b}
g^{(2)}(0)=\frac{\langle\hat{a}_{1}^{\dagger2}\hat{a}_{1}^{2}\rangle}{\langle\hat{a}_{1}^{\dagger}\hat{a}_{1}\rangle^{2}}\simeq\frac{\eta_{1}^{2}\left(\Delta_{4}\Delta_{6}/\sigma_{2}+J^{2}\chi\right)^{2}}{\eta_{2}^{2}\sigma_{2}^{2}\Delta_{4}^{4}}.
\end{equation}

The correlation function can be calculated numerically by solving the quantum master equation~\cite{johansson2013QuTiP,johansson2012QuTiP}
\begin{equation}\label{eq:master}
\dot{\hat{\rho}}=i[\hat{\rho}, \hat{H}_2]+\sum_{j=1,2}\frac{\gamma}{2}\left(2 \hat{a}_j \hat{\rho} \hat{a}^{\dagger}_j-\hat{a}^{\dagger}_j \hat{a}_j \hat{\rho}-\hat{\rho} \hat{a}^{\dagger}_j \hat{a}_j\right).
\end{equation}
The photon-number probability is $P_{mn}=\langle m,n |\hat{\rho}_\text{ss}|mn\rangle$, which can be obtained from the steady-state solutions $\hat{\rho}_\text{ss}$ of the master equation. An excellent agreement between the analytical and numerical results is seen in Figs.~\ref{Fig2}(c) and \ref{Fig2}(d).

Figure~\ref{Fig2} shows quantum and classical nonreciprocities with backscattering. In Fig.~\ref{Fig2}(b), for the resonant cases, we find that the switch between CNR and QNR can still be achieved by tuning the angular velocity $\Omega$ after considering the effect of backscattering. When $\Omega$ from $0$ to $6\,\mathrm{kHz}$, the classical nonreciprocity $(\mathcal{R}_{1}\neq0)$ appears, but there is no quantum nonreciprocity $(\mathcal{R}_{2}=0)$.  Adding $\Omega$ beyond $17\,\mathrm{kHz}$, the opposite occurs.

Different from the above resonant cases, Fig.~\ref{Fig2}(c) shows that no classical nonreciprocity occurs at $\Delta_{L}=-0.98\,\mathrm{kHz}$ by fixing the rotational speed $\Omega=3.8$\, kHz. In addition, for $J/\gamma=2$, the number of peaks of $N$ increased from one to two, compared with an ideal resonator [Fig.~\ref{Fig1}]. A similar feature can also be identified in the number of dips of $g^{(2)}(0)$ in Fig.~\ref{Fig2}(d). The reason is that the energy level splitting caused by backscattering provides the possibility of more photon jumps. This means we can explore the richer significance of nonreciprocity by combining backscattering and mechanical rotation.

As shown in Fig.~\ref{Fig3}(a), for $J/\gamma<0.7$, the second-order correlation functions of the two modes are reciprocal ($\mathcal{R}_2=0$) for smaller angular velocity $\Omega<1.9\,\mathrm{kHz}$, which is because the mode splitting due to backscattering and the Sagnac-Fizeau shift due to rotation are both small. When $J/\gamma>0.7$, the minimum angular velocity required for the emergence of quantum nonreciprocity ($\mathcal{R}_2\neq0$) increases with increasing backscattering rate. Also, we find that various kinds of nonreciprocity can be switched by tuning the backscattering and mechanical rotation of the resonator for the resonant cases [Fig.~\ref{Fig3}].

\begin{figure*}
\centering
\includegraphics[width=1.02\textwidth]{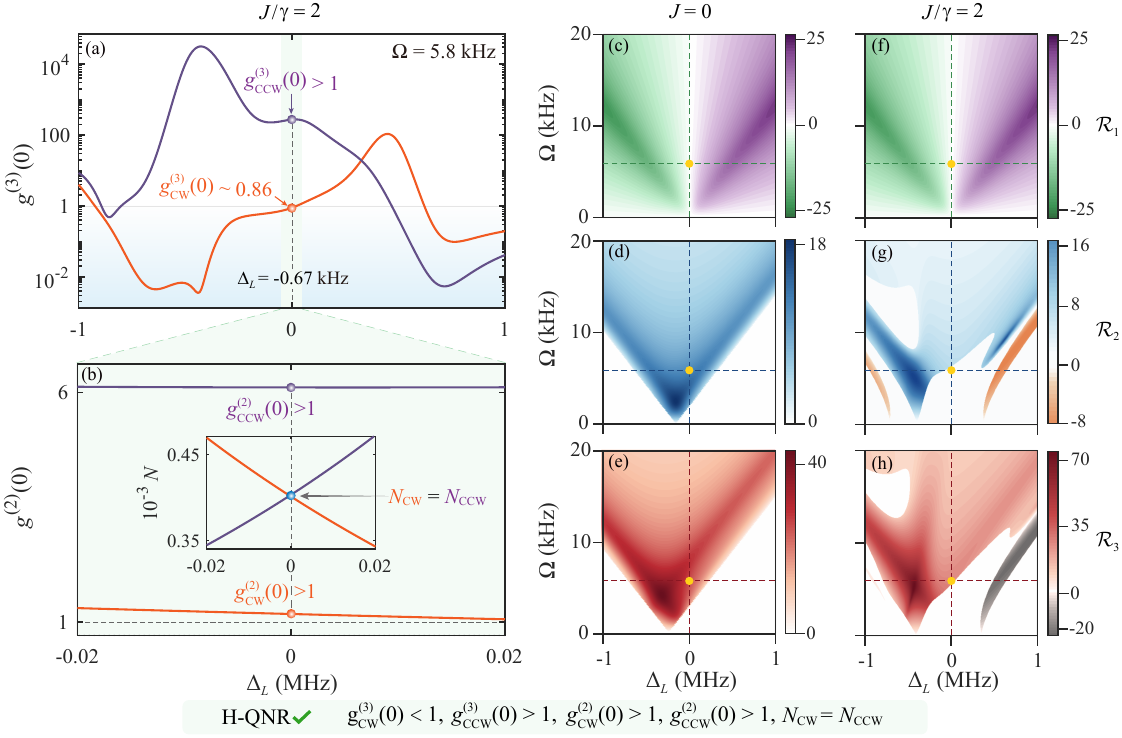}
\caption{The backscattering-induced higher-order quantum nonreciprocity. (a, b) The correlation functions $g^{(3)}(0)$, $g^{(2)}(0)$ and the mean photon number $N$ versus the optical detuning $\Delta_{L}$ for $J/\gamma=2$ and $\Omega=5.8~\mathrm{kHz}$ (i.e., $\Omega/\gamma=0.024$). (c)-(h) For the generation of higher-order quantum nonreciprocity, the comparison between the cases with and without backscattering. The other parameters are the same as those in Fig.~\ref{Fig1}.}\label{Fig4}
\end{figure*}
Further, we extend our research to the higher-order quantum nonreciprocity of this system. The third-order correlation function can be obtained analytically as
\begin{equation}\label{eq:g3b}
g^{(3)}(0)=\frac{\langle\hat{a}_{1}^{\dagger3}\hat{a}_{1}^{3}\rangle}{\langle\hat{a}_{1}^{\dagger}\hat{a}_{1}\rangle^{3}}\simeq\frac{\eta_{1}^{4}\left(\eta_{3}\Gamma_{1}-\Delta_{8}\Gamma_{2}\right)^{2}}{\mu^{2}\eta_{2}^{2}\sigma_{2}^{2}\Delta_{4}^{6}}.
\end{equation}
The condition $g^{(3)}(0)<1$ [$g^{(3)}(0)>1$] indicates the third-order
sub-Poissonian (super-Poissonian) statistics, which is also
interpreted as three-photon antibunching (bunching) in recent
experiments on multi-photon blockade~\cite{hamsen2017two} and
photon-induced tunneling~\cite{Rundquist14}. Similar to Eq.~(\ref{eq:R}), we define a higher-order quantum nonreciprocity ratio $\mathcal{R}_{3}$, which are written as
\begin{equation}\label{eq:R3}
\mathcal{R}_{3}= 10\log_{10}\frac{g_{\mathrm{CCW}}^{(3)}(0)}{g_{\mathrm{CW}}^{(3)}(0)}.
\end{equation}
We set $\mathcal{R}_{3}=0$ for the cases of both
$g_\mathrm{CCW}^{(3)}(0)$ and $g_\mathrm{CW}^{(3)}(0)$ are smaller or larger than 1. The difference is that the condition of higher-order quantum nonreciprocity (H-QNR) not only needs to satisfy $\mathcal{R}_{3}\neq 0$, but also needs to satisfy the condition of reciprocity of classical mean photon number ($\mathcal{R}_{1}=0$) and quantum second-order correlation function ($\mathcal{R}_{2}=0$).

\vspace{3mm}
Figure~\ref{Fig3}(b) shows the relationship of higher-order quantum nonreciprocity with backscattering and angular velocity for the resonant cases. We find that higher-order quantum nonreciprocity occurs as the backscattering rate $J$ increases beyond a certain point. For $J/\gamma=2$ [Fig.~\ref{Fig3}(b-i)], higher-order quantum nonreciprocity emerges with angular velocity $5.56\,\mathrm{kHz}\geq\Omega\leq6.17\,\mathrm{kHz}$. This result is in agreement with the results of Figs.\ref{Fig4}(a) and \ref{Fig4}(b). Increasing $J/\gamma$ to $2.9$, the range of angular velocity $\Omega$ corresponding to the appearance of higher-order quantum nonreciprocity is $10.13-11.28\,\mathrm{kHz}$ [Fig.~\ref{Fig3}(b-ii)]. By further the adding backscattering rate ($J/\gamma=3.8$), the corresponding $\Omega$ is increased to $16.59-17.74\,\mathrm{kHz}$ [Fig.~\ref{Fig3}(b-iii)]. It shows that the minimum angular velocity $\Omega$ required for the emergence of higher-order quantum nonreciprocity increases with increasing backscattering rate $J$. We also find that higher-order quantum nonreciprocity can be achieved by controlling the angular velocity and the backscattering rate of the resonator for the resonant cases [Fig.~\ref{Fig3}(b)].

\vspace{3mm}
Figures~\ref{Fig4}(a) and~\ref{Fig4}(b) show that higher-order quantum nonreciprocity occurs around $\Delta_{L}=-0.67\,\mathrm{kHz}$. At this point, for different input directions, the third-order correlation functions are nonreciprocal ($g_{\mathrm{CW}}^{(3)}(0)\sim0.86$, $g_{\mathrm{CCW}}^{(3)}(0)\sim272.68$), but the second-order correlation functions and the mean photon numbers are reciprocal, i.e., $g_{\mathrm{CW}}^{(2)}(0)>1$, $g_{\mathrm{CCW}}^{(2)}(0)>1$, and $N_{\mathrm{CW}}=N_{\mathrm{CCW}}$. As far as we know, this pure higher-order quantum nonreciprocity has not been revealed in previous works on the nonreciprocal effect.

To clearly show the difference in nonreciprocities between the ideal cavity ($J=0$) and the realistic cavity $(J/\gamma=2)$, we compare the nonreciprocity ratios of the above different cases, as shown in Figs.~\ref{Fig4}(c-h). We find that backscattering has a small effect on the classical nonreciprocity [see Figs.~\ref{Fig4}(c) and \ref{Fig4}(f)], but a large effect on the quantum nonreciprocity, e.g., the negative value of $\mathcal{R}_{3}$ appears in Fig.~\ref{Fig4}(h). By comparison, we note that the higher-order quantum nonreciprocity occurs in the realistic cavity, due to the effect of backscattering, but not in an ideal cavity.

\vspace{5mm}
\section{Conclusions\label{sec:5}}
In summary, we study the coherent switch of classical and quantum nonreciprocities of photons, and realize the higher-order quantum nonreciprocity by using a spinning Kerr resonator. Our findings contain three main features. First, we show a single device switching between a classical isolator and a purely quantum directional system by adjusting multiple degrees of freedom (the optical detuning $\Delta_{L}$ and angular velocity $\Omega$). Furthermore, we present the higher-order quantum nonreciprocity for the first time, which provides a richer degree of freedom for one-way optical control, i.e., it is available to achieve one-way quantum communication while classical communication is reciprocal. More interestingly, in practical devices, the backscattering is unavoidable, but it can induce higher-order quantum nonreciprocity. These results offer the possibility of new developments in nonreciprocal devices, as well as have potential applications in nanoparticle sensing. We believe that our work can be extended to spinning photonics and spinning optomechanics with similar backscattering.
\vspace{5mm}
\section{ACKNOWLEDGMENTS}
H.J. is supported by the National Natural Science Foundation of China (Grants No.~11935006 and No.~11774086) and the Science and Technology Innovation Program of Hunan Province (Grant No.~2020RC4047). R.H. is supported by the Japan Society for the Promotion of Science (JSPS) Postdoctoral Fellowships for Research in Japan (No.~P22018). X.-W.X. is supported by the National Natural Science Foundation of China (NSFC) (Grants No.~12064010 and  No.~12247105), Natural Science Foundation of Hunan Province of China (Grant No.~2021JJ20036), and the science and technology innovation Program of Hunan Province (Grant No.~2022RC1203).

\end{document}